\documentstyle[epsfig,12pt]{article}

\input amssym.def       
\input amssym.tex

\def\double{\Bbb}

\def\cc{{\double C}}     
\def\nn{{\double N}}       
\def\zz{{\double Z}}
\def\rr{{\double R}}     
\def\kk{{\double K}}

\def\aa{{\cal A}}
\def\bb{{\cal B}}
\def\ccc{{\cal C}}
\def\dd{{\cal D}}
\def\ee{{\cal E}}
\def\gg{{\cal G}}        
\def\hh{{\cal H}}
\def\hhh{{\double H}}   
\def\mm{{{\cal M}}}    
\def\nnn{{\cal N}}
\def\jj{{\cal J}}
 
\def\t{\mathrm{Tr}} 
\def\tp{\mathrm{Tr_p}}

\def\lb{\left[} 
\def\rb{\right]}
\def\lp{\left(} 
\def\rp{\right)}
\def\la{\left\{} 
\def\ra{\right\}}

\def\ul{\underline}
\def\ov{\overline}
\def\ot{\otimes}
\def\op{\oplus}
\def\om{\ominus}

\def\bbb{\begin{eqnarray}}
\def\eee{\end{eqnarray}}
\def\pp{\pmatrix}
\def\n{\nonumber}

\begin{document}

\hsize 17truecm
\vsize 24truecm
\font\twelve=cmbx10 at 13pt
\font\eightrm=cmr8
\baselineskip 16pt

\begin{titlepage}

\centerline{\twelve CENTRE DE PHYSIQUE THEORIQUE}
\centerline{\twelve CNRS - Luminy, Case 907}
\centerline{\twelve 13288 Marseille Cedex 9}
\vskip 2truecm
\begin{center}
{\bf\Large  \sc CLASSIFICATION OF \\
\medskip
FINITE SPECTRAL TRIPLES}
\end{center}
\bigskip
\begin{center}
{\bf Thomas KRAJEWSKI}
\footnote{ and Universit\'e de Provence and Ecole Normale Superieure de Lyon, tkrajews@cpt.univ-mrs.fr} \\
\end{center}
\bigskip
\vskip 2truecm
\leftskip=1cm
\rightskip=1cm
\centerline{\bf Abstract} 
\bigskip
It is known that the spin structure on a Riemannian manifold can be extended to noncommutative geometry using the notion of a spectral triple. For finite geometries, the corresponding finite spectral triples are completely described in terms of matrices and classified using diagrams. When tensorized with the ordinary space-time geometry, finite spectral triples give rise to Yang-Mills theories with spontaneous symmetry breaking, whose characteristic features are given within the diagrammatic approach: vertices of the diagram correspond to gauge multiplets of chiral fermions and links to Yukawa couplings.  
\vskip 1truecm
PACS-92: 11.15 Gauge field theories\\ 
\indent
MSC-91: 81T13 Yang-Mills and other gauge theories\\ 
\vskip 1truecm

\noindent December 1996\\
\vskip 0.2truecm
\noindent
CPT-96/P.3409\\
\end{titlepage}

\section{Introduction}

It is nowadays admitted that the major difficulty in quantizing gravity lies in our current conception of geometrical spaces at very short scale. Therefore, a first step towards a quantum theory of gravitation may be a suitable redefinition of the geometrical notions. Due to Heisenberg's uncertainty relations, in a quantum theory, the ordinary notion of a point disappears, so that we have to develop a geometry without any reference to points. This is the basic principle of noncommutative geometry, where points are replaced by the algebra of  coordinates, which may be noncommutative. There are many possible choices of the algebra of coordinates. For example, we can replace the ordinary plane, whose coordinates satisfy $xy=yx$, by the quantum plane, i.e. the algebra generated by the relation $xy=qyx$ for a complex number $q$. This leads us naturally, when we aim at describing the symmetries of the quantum plane, to the theory of quantum groups.
\par
Another possible choice of the algebra of coordinates lies in replacing it by a suitable involutive subalgebra of the algebra of operators on a Hilbert space. Then, we try to extend, guided by the analogy with quantum mechanics, the ordinary definitions and theorems of differential geometry, to the noncommutative case. This is the framework of {\it noncommutative geometry}, a theory fully developped in \cite{connes1}-\cite{madore}. Since matter is fermionic, the notion that should be extended to noncommutative geometry if that of a spin structure on a riemannian manifold. This has been achieved via {\it spectral triples} in \cite{connes4},\cite{connes5}. By definition, a spectral triple is a triple $(\aa,\hh,\dd)$, where $\aa$ is a real involutive algebra, that stands for the algebra of coordinates, represented on the fermionic Hilbert space $\hh$. $\dd$ is a self-adjoint operator acting on $\hh$, which is the  generalisation of the Dirac operator. To provide a suitable extension of spin geometry, these three objects are assumed to fulfill some axioms called axioms of noncommutative geometry. 
\par
Starting with a suitable class of spectral triples, we can construct gauge theories with spontaneous symmetry breaking. In that case, both the gauge and the Higgs fields appear as connections, the former on an ordinary bundle, the latter on a finite one. Accordingly, the spectral triples we use are products of two spectral triples. The first one encodes all data concerning the geometry of spacetime and the second one is what we call a {\it finite spectral triple}.      \par
In section 2, we define finite spectral triples and give the relation between the general axioms of noncommutative geometry and the simplest case of finite noncommutative geometry. Section 3 is devoted to a general resolution of the equations we obtained. In section 4, we introduce some diagrams that allow to classify and to construct finite spectral triples. Finally, in section 5 we give some applications to model building in particle physics and determine the general feature of the corresponding Yang-Mills-Theory within the diagrammatic approach.

\section{Axioms for finite spectral triples}

Finite spectral triples are particular cases of spectral triples of dimension 0. The latter are rigorously defined within the axioms of noncommutative geometry and yield a general theory of discrete spaces. Among all discrete spaces, we focus on finite ones, thus the algebra is finite dimensional.  Futhermore, we will also assume that the Hilbert space is finite dimensional, an infinite dimensional one corresponding to a theory with an infinite number of elementary fermions. Accordingly, a finite spectral triple $(\aa,\hh,\dd)$ is defined as a spectral triple of dimension 0 such that both $\aa$ and $\hh$ are finite dimensional. Using such a triple, it is possible to a construct Yang-Mills theory with spontaneous symmetry breaking whose gauge group is the group of unitary elements of  $\aa$, $\hh$ is the fermionic Hilbert space and $\dd$ is the mass matrix.
\par
It is  known that finite dimensional real involutive algebras which admit a faithful representation on a finite dimensional Hilbert space are just direct sums of matrix algeras over the fields of real numbers, complex numbers and quaternions. Therefore, we write the algebra as a direct sum
$$
\aa=\mathop{\op}\limits_{i=1}^{N}M_{n_{1}}(\kk),
$$
where $M_{n}(\kk)$ denotes the algebra of square matrices of order $n$ with entries in the field $\kk=\rr$, $\cc$ or $\hhh$. This remark simplifies considerably finite noncommutative geometry and it becomes possible to give a detailed account of all finite spectral triples. We also consider the particular case of complex finite spectral triples, whose algebra is a complex algebra and the representation $\pi$ is assumed to be linear over complex numbers.
\par  
In this section, we give the axioms for finite spectral triples and sketch their relations with the general case. Since both the algebra $\aa$ and the Hilbert space $\hh$ are finite dimensional, all assumptions concerning functional analysis, that are necessary in the general case, are obviously satisfied. Thus, there is a overwhelming simplification in the formulation of noncommutative geometry and the seven axioms involved in \cite{connes5} reduce to the four following ones: {\it reality, orientability, Poincar\'e duality} and {\it first order condition}. To proceed, we fix a finite spectral triple $\lp\aa,\hh,\dd\rp$ together with a faithful and unitary representation $\pi$ of $\aa$ as operators on $\hh$. Moreover, the formulation of the axioms of noncommutative geometry involves two other operators, the {\it charge conjugation} $\jj$ and the {\it chirality} $\chi$ that we shall also fix. The precise definition of these two operators will become clearer in the following sections. 

\subsection{Reality}

To define noncommutative manifolds, we need, in this setting, a suitable notion of Poincar\'e duality. A first step towards a formulation of Poincar\'e duality in noncommutative geometry lies in introducing a bimodule structure on $\hh$. Note that the bimodule structure is also necessary if one wants to incorporate the color sector in the Connes-Lott model \cite{connes3}. However, this bimodule is not arbitrary and it has to fulfill the reality axiom.
\vskip 0.5truecm

\ul{{\bf Reality (finite case)}}
\vskip 0.2cm
\noindent
{\it There is an antiunitary involution $\jj$ on $\hh $(i.e. $\jj^{*}=\jj^{-1}=\jj$) that commutes with $\chi$ 
and such that, for any $x,y\in\aa$ one has $\lb\pi(x),\jj\pi(y)\jj^{-1}\rb=0$.}\\ 
\vskip 0.1truecm
Then we can equip $\hh$ with a bimodule structure. The left action is
$$
(x,\psi)\in\aa\times\hh\mapsto x\psi=\pi(x)\psi,
$$
and the right action is
$$
(\psi,y)\in\hh\times\aa\mapsto\jj\pi(y^{*})\jj^{-1}\psi.
$$
This defines an $(\aa,\aa)$-bimodule structure or equivalently, a $(\aa\ot\aa^{op})$-module structure on $\hh$. Note that at a mathematical level, the introduction of this operator has been motivated by the transition from complex K-theory to real K-theory that turns out to be necessary in the formulation of noncommutative spin geometry. Moreover, given a cyclic and separating vector in $\hh$, $\jj$ is Tomita-Takesaki involution that allows us to pass from $\aa$ to its commutant $\aa'$.\\
For the standard model, this operator is just charge conjugation. In other words, the Hilbert space of the finite spectral triple of the standard model is the vector space spanned by all fermions and antifermions and the operator $\jj$ is the antilinear operator mapping a fermion to the corresponding antifermion.
\par
We gave the reality axiom in the finite case only. For general spectral triples, we only have to change some signs in the relations involving $\jj$, depending on the {\it dimension} modulo 8 of the spectral triple \cite{connes5}. In the commutative case of a smooth manifold $V$ endowed with a spin structure, $\jj$ is the charge conjugation defined on the space of square integrable sections of the spinor bundle of $V$.   

\subsection{Orientability}

Since the noncommutative manifold defined by a spectral triple has to be oriented, we must specify an orientation form. In the case of an ordinary manifold of dimension n, this is achieved by means of a nowhere vanishing n-form $\gamma$. The extension of this notion to the noncommutative case goes through the deep relation, in the commutative case, between differential forms and Hochschild homology. 
\par
Let us now briefly recall some elementary aspects of Hochschild homology, referring to \cite{loday} for a general theory. Given an associative, unital, but not necessarily commutative algebra $\aa$, and a bimodule $\mm$ over $\aa$, we first define, for any integer n, the vector space $T_{n}(\aa,\mm)$ as the tensor product of $\mm$ by n copies of $\aa$
$$
T_{n}(\aa,\mm)=\mm\ot\aa\ot...\ot\aa.
$$    
Then, we define, for any $n>0$, the boundary operation $b_{n}$ as a linear map from $T_{n}(\aa,\mm)$ to $T_{n-1}(\aa,\mm)$ such that  
\bbb
&b_{n}(m\ot x_{1}\ot...\ot x_{n})=
mx_{1}\ot x_{2}\ot...\ot x_{n}+(-1)^{n}x_{n}m\ot x_{1}\ot ...\ot x_{n-1}&\n\\
&+\mathop{\sum}\limits_{i=1}^{n-1} (-1)^{i} m\ot x_{1}\ot...\ot x_{i}x_{i+1}\ot...\ot x_{n}.\n&
\eee
For $n=0$ we set $b_{0}=0$. By definition, the Hochschild cycles $Z_{n}(\aa,\mm)$ of dimension n are the elements of $T_{n}(\aa,\mm)$ whose image under $b_{n}$ is zero. Now, let us consider the bimodule $\mm=\aa\ot\aa^{op}$, whose bimodule structure is given by 
$$
y(x\ot x')y'=yxy'\ot x',
$$
for any $y,y'\in\aa$ and $x\ot x'\in\aa\ot\aa^{op}$.
\par
These rather abstract objects are represented as operators on the Hilbert space. Using the Dirac operator $\dd$, the representation $\pi$ can be extended to the Hochschild cycles as
$$
\pi(x\ot x'\ot y_{1}\ot...\ot y_{n})=\pi(x)\jj\pi(x')\jj^{-1}
\lb\dd,\pi(y_{1})\rb...\lb\dd,\pi(y_{n})\rb,
$$ 
for any $x\ot x'\ot y_{1}\ot...\ot y_{n}\in Z_{n}(\aa,\aa\ot\aa^{op})$. This leads to a natural extension of the orientation of a manifold in the noncommutative case \cite{connes5}. 
\vskip 0.5truecm
\ul{{\bf Orientability}}
\vskip 0.2cm
\noindent
{\it For a noncommutative geometry of dimension n (n even), the chirality $\chi$ is a hermitian involution that commutes with $\pi(x)$ for any $x\in\aa$, anticommutes with the Dirac operator $\dd$ and is the image of an n-dimensional Hochschild cycle}.\\
\vskip 0.1truecm
Before we focus on finite spectral triples, let us check that it holds for the ordinary euclidian geometry of spacetime. The latter is a compact four dimensional manifold $V$, the algebra $\aa$ is the algebra $C^{\infty}(V)$ of complex valued smooth functions on $V$, the Hilbert space $\hh$ is the space $L^{2}(V,S)$ of square integrable sections of the spinor bundle of $V$, $\dd$ is the ordinary Dirac operator defined in a local chart by $\dd=i\gamma^{\mu}\partial_{\mu}$, $\gamma^{\mu}$ being the usual euclidian Dirac matrices, and $\jj$ is the charge conjugation of Dirac spinors. The functions act simply on $\hh$ by multiplication and we have, for any $f\in C^{\infty}(V)$,
$$
\lb\dd,\pi(f)\rb=i\gamma^{\mu}\partial_{\mu}f.
$$ 
Since the algebra is commutative, the bimodule $\mm$ involved in the definition of Hochschild homology is not relevant and we simply take the Hochschild cycles $Z_{n}(\aa,\mm)$ with $\mm=\aa$. Accordingly, any 0-form is a function and can be written as the image of a 0 dimensional cycle. Moreover, Hochschild cycles of dimension 1 give us 1-forms. Indeed, starting with the cycle $\sum_{i} f_{0}^{i}\ot f_{1}^{i}$, its image is  
$$
\sum_{i}if_{0}^{i}\partial_{\mu}f_{1}^{i}\gamma^{\mu},
$$ 
that can be considered as the 1-form $A_{\mu}dx^{\mu}$ if we identify $\sum_{i}f_{0}^{i}\partial_{\mu}f_{1}^{i}$ with $A_{\mu}$ and $i\gamma^{\mu}$ with $dx^{\mu}$.
\par
At higher orders, the situation is more subtle but we can still construct, using antisymmetrisation, some Hochschild cycles whose images are  differentials forms if we identify $i\gamma^{\mu}$ with $dx^{\mu}$. In fact, for any $f_{0},f_{1},...,f_{n}\in C^{\infty}(V)$, we define
$$
c=\mathop{\sum}\limits_{\sigma}\epsilon(\sigma)f_{0}\ot f_{\sigma(1)}\ot...\ot f_{\sigma(n)},
$$  
where the sum runs over all permutations of ${1,2,..,n}$ and $\epsilon(\sigma)$ is the signature of $\sigma$. Due to the commutativity of $C^{\infty}(V)$, we  check that $c$ is a Hochschild cycle and that, thanks to the antisymmetrisation procedure, its image is a n-form. This applies for the volume form that can be defined as
$$
dx^{0}\wedge dx^{1}\wedge dx^{2}\wedge dx^{3}
=\frac{1}{4!}\epsilon_{\mu\nu\rho\sigma}
dx^{\mu}\wedge dx^{\nu}\wedge dx^{\rho}\wedge dx^{\sigma}.
$$ 
If we identify $dx^{\mu}$ with $i\gamma^{\mu}$, the previous relation is the definition of $\gamma^{5}$ so that $\gamma^{5}$ turns out to be the chirality, and it corresponds to the volume form.
\par
Let us  give two remarks pertaining to the commutative case. First, the antisymmetrisation procedure yields us immediately differential forms, so that we do not have to divide by some auxiliary fields (the so-called "junk" defined in \cite{connes3}). Secondly, we have seen that Hochschild cycles provide us with differential forms when represented as operators on the Hilbert space. Thus, it is natural to ask for an exterior derivative in this framework. More precisely, we would like to have some linear maps $d_{n}$ from $Z_{n}(\aa,\aa)$ into $Z_{n+1}(\aa,\aa)$ such that $d_{n+1}\,\circ\, d_{n}=0$ and such that, when represented with Dirac matrices, they coincide with the genuine exterior derivatives. In fact, we can set $d_{n}=B_{n}$, where $B_{n}$ is Connes' coboundary map \cite{loday}. For an associative and unital algebra $\aa$, the latter is defined as a linear map from $T_{n}(\aa,\aa)$ into $T_{n+1}(\aa,\aa)$ by
\bbb
&B_{n}(x_{0}\ot x_{1}\ot...\ot x_{n})=&\n\\
&\mathop{\sum}\limits_{i=0}^{n-1}(-1)^{ni}\,
1\ot x_{i}\ot...\ot x_{n}\ot x_{0}\ot...\ot x_{i-1}&\n\\
&-\mathop{\sum}\limits_{i=0}^{n-1}(-1)^{n(i-1)}\,
x_{i-1}\ot 1\ot x_{i}\ot...\ot x_{n}\ot x_{0}\ot...\ot x_{i-2},&\n
\eee
for any $x_{0},...,x_{n}\in\aa$. It is easy to check that $B_{n-1}\,\circ\,b_{n}+b_{n+1}\,\circ\,B_{n}=0$. Therefore, $B_{n}(Z_{n}(\aa,\aa))$ is included in $Z_{n+1}(\aa,\aa)$. Moreover, one has $B_{n+1}\,\circ\, B_{n}=0$ and thus $B_{n}$ can be considered as an exterior derivative in the framework of Hochschild homology. Besides, in the commutative case, when we lift this coboundary operation on operators on the Hilbert space using Dirac matrices, it yields the genuine exterior derivative without intoducing auxiliary fields. Let us check it for 1-forms. If $A_{\mu}dx^{\mu}$ is a 1-form, it can be written as the image of the Hochschild cycle $\sum_{i} f_{0}^{i}\ot f_{1}^{i}$, with $A_{\mu}=\sum_{i}f_{0}^{i}\partial_{\mu}f_{1}^{i}$. The action of $B_{1}$ on this cycle is given by
\bbb
&B_{1}\lp\sum_{i}f_{0}^{i}\ot f_{1}^{i}\rp=&\n\\
&\mathop{\sum}\limits_{i}\lp 1\ot f_{0}^{i}\ot f_{1}^{i}-1\ot f_{1}^{i}\ot f_{0}^{i}
+f_{0}^{i}\ot 1\ot f_{1}^{i}- f_{1}^{i}\ot 1\ot f_{0}^{i}\rp.&\n
\eee
When we represent it, we end up with
$$
i\gamma_{\mu}i\gamma_{\nu}\lp
\partial_{\mu} f_{0}^{i}\partial_{\nu} f_{1}^{i}
-\partial_{\mu} f_{1}^{i}\partial_{\nu} f_{0}^{i}\rp=i\gamma_{\mu}i\gamma_{\nu}\lp
\partial_{\mu}A_{\nu}-\partial_{\mu}A_{\nu}\rp=F_{\mu\nu}dx^{\mu}\wedge dx^{\nu},
$$
with $F_{\mu\nu}=\partial_{\mu}A_{\nu}-\partial_{\mu}A_{\nu}$ and if we identify $i\gamma^{\mu}$ with $dx^{\mu}$. Let us note that this also works for higher order forms. 
\par
For finite spectral triples, the orientability axiom requires that $\chi$ must be written as the image of a Hochschild cycle of dimension 0. Since all elements of $T_{0}(\aa,\aa\ot\aa^{op})$ are cycles, it only states that there are some elements $a_{i},b_{i}\in\aa$ such that $\chi=\sum_{i}\pi(a_{i})\jj\pi(b_{i})\jj^{-1}$. Accordingly, the orientability axiom is reformulated as follows.
\vskip 0.5truecm
\ul{{\bf Orientability (finite case)}}
\vskip 0.2cm
\noindent
{\it The chirality $\chi$ is an hermitian involution that commutes with $\pi(x)$ for any $x\in\aa$, anticommutes with $\dd$ and can be written as $\chi=\sum_{i}\pi(x_{i})\jj\pi(y_{i})\jj^{-1}$ for some $x_{i},y_{i}\in\aa$.}\\
\vskip 0.1truecm
For the standard model, $\chi$ is simply the operator of $\hh$ that takes the value $-1$ for left-handed fermions and $+1$ for right-handed fermions.   

\subsection{Poincar\'e duality}

For a closed Riemannian manifold $V$ of dimension $n$, there is a well known isomorphism between the de Rham groups $H^{p}(V)$ and $H^{n-p}(V)$ given by the bilinear map
$$
(f,g)\mapsto\int_{V} f\wedge *g,
$$ 
where $f$ and $g$ are closed p and (n-p)-forms that are representatives of the de Rham groups $H^{p}(V)$ and $H^{n-p}(V)$. Formulating this in noncommutative geometry is a rather difficult task that requires the use of Kasparov's biinvariant theory. We refer to \cite{connes2} and $\cite{connes5}$ for a detailed account of Poincar\'e duality in noncommutative geometry. In this section, we simply give the axiom of Poincar\'e duality, as stated in \cite{connes5}, and derive its formulation in the finite case. The Poincar\'e duality axiom is
\vskip 0.5truecm
\ul{{\bf Poincar\'e duality}}
\vskip 0.2cm
\noindent
{\it The intersection form $K_{*}(\aa)\times K_{*}(\aa)\rightarrow\zz$ is non degenerate.}\\
\vskip 0.1truecm
Some insights on $V$ are given by the knowledge of the vector bundles over $V$ through its K-theory that can be formulated algebraically. $K_{*}(\aa)$ is a collective notation for all K-groups. Due to Bott periodicity, the sequence of these groups is periodic with period 2, for complex algebras. In the case of real algebras, the period is 8 \cite{schroeder}. However, if we do not take into account the torsion (i.e. the presence of K-groups of type $\zz/2\zz$), only matters the group $K_{0}(\aa)$. Since $K_{0}$ is additive, that is,
$$
K_{0}\lp\mathop{\op}\limits_{i=1}^{N}M_{n_{i}}(\kk)\rp=
\mathop{\op}\limits_{i=1}^{N}K_{0}\lp M_{n_{i}}(\kk)\rp,
$$
we have to compute $K_{0}(M_{n}(\kk))$ \cite{brodzki}. By definition, $K_{0}(M_{n}(\kk))$ is the Grothendieck group of stably isomorphic classes of finitely generated projective modules over $\aa$. Recall that a module over $M_{n}(\kk)$ is a vector space $E$ equipped with a representation of $M_{n}(\kk)$ which is said to be finitely generated if any of its elements can be written as a linear combination with coefficients in $M_{n}(\kk)$ of a finite number of elements of $E$. The additive structure is given by the direct sum of modules, and two such modules are said to be stably isomorphic if they are isomorphic after addition of a trivial module (i.e. a module that is simply a direct sum of several copies of the algebra).
\par
The direct sum of modules yielding only an addition, we have to define the opposite of a module to obtain a group. This is achieved by means of the Grothendick procedure, which is used to construct the additive group of integers $\zz$ out of the positive integers $\nn$. Thus, we end up with an abelian group, which turns out to be the free group generated by the class of the "smallest" modules over $M_{n}(\kk)$. The latter is given by any non zero projection in $M_{n}(\kk)$ of minimal rank, since two projections of same rank are obviously equivalent and since the rank of the direct sum of two projections is the sum of their ranks. Now, we simply have to find such a projection in the real, complex and quaternionic cases. In the first two cases, the solution is obvious since any matrix whose only non vanishing entry on the diagonal is equal to 1 is a projection of minimal rank. Therefore such a projection has trace 1. In the quaternionic case, it remains true if we consider that the only non vanishing entry is the unity of $\hhh$, so that it has trace 2. Finally, $K_{0}(\aa)$ turns out to be $\zz^{N}$ and is the free group generated by $(p_{i})_{1\leq i\leq N}$, where $p_{i}\in M_{n_{i}}(\kk)$ is a self-adjoint projection of minimal rank. As an additive group, it can be considered as a $\zz$-module with basis $(p_{i})_{1\leq i\leq N}$.
\par
To formulate Poincar\'e duality for finite spectral triples, we have to compute the intersection form $\cap$ in this case. More precisely, since $\cap$ is a $\zz$-bilinear form on $K_{0}(\aa)\times K_{0}(\aa)$, we have to compute its matrix given by $\cap_{ij}=\cap(p_{i},p_{j})$. To proceed, we apply the the general definition of the intersection form as the pairing of the Dirac operator index with the two projections $p_{i}$ and $p_{j}$. Finite spectral triples are even spectral triples, so that we rewrite the Dirac operator as
$$
\dd=M+M^{*},
$$ 
where
$$
M=1/4(1-\chi)\,\dd\, (1+\chi).
$$  
$M$ is a map from the space of right-handed fermions (eigenspace of $\chi$ associated to the eigenvalue +1) to the space of left-handed fermions (eigenspace of $\chi$ associated to the eigenvalue -1), and can be considered as a mass matrix. Using the reality axiom, the Hilbert space is endowed with a $\aa\ot\aa^{op}$-module structure that allows us to construct the linear map
$$
M_{ij}=\pi(p_{i})\jj\pi(p_{j})\jj^{-1}\, M\,\pi(p_{i})\jj\pi(p_{j})\jj^{-1}.
$$
In fact, $M_{ij}$ is a map from $\hh_{ij}^{R}$ into $H_{ij}^{L}$, where 
\bbb
\hh_{ij}^{R}&=&1/2\pi(p_{i})\jj\pi(p_{j})\jj^{-1}(1+\chi)\hh,\n\\
\hh_{ij}^{L}&=&1/2\pi(p_{i})\jj\pi(p_{j})\jj^{-1}(1-\chi)\hh.\n
\eee
Both spaces being finite dimensional, the index is easily computed and we get,
$$
\mathrm{Index}(M_{ij})=\mathrm{dim}(\hh_{ij}^{R})-\mathrm{dim}(\hh_{ij}^{L}).
$$
Since $\chi$ commutes with $\pi(p_{i})$ and $\jj\pi(p_{j})\jj^{-1}$, the operators $1/2\pi(p_{i})\jj\pi(p_{j})\jj^{-1}(1+\chi)$ and $1/2\pi(p_{i})\jj\pi(p_{j})\jj^{-1}(1-\chi)$ are projections. Therefore, the dimension of $\hh_{ij}^{R}$ and $\hh_{ij}^{L}$ are simply the trace of the corresponding projections. Finally, $\cap_{ij}$, defined as the pairing of $p_{i}\ot p_{j}$ with the index of $\dd$, is given by
\bbb
\cap_{ij}&=&\mathrm{Index}(M_{ij})\n\\
&=&\mathrm{dim}(\hh_{ij}^{R})-\mathrm{dim}(\hh_{ij}^{L})\n\\
&=&1/2\t\lb\pi(p_{i})\jj\pi(p_{j})\jj^{-1}(1+\chi)\rb-
1/2\t\lb\pi(p_{i})\jj\pi(p_{j})\jj^{-1}(1-\chi)\rb\n\\
&=&\t\lb\chi\lp\pi(p_{i})\jj\pi(p_{j})\jj^{-1}\rp\rb.\n
\eee
Accordingly, we have obtained a simple formulation of Poincar\'e duality for the finite spectral triples that can be stated as follows.
\vskip 0.5truecm
\ul{{\bf Poincar\'e duality (finite case)}}
\vskip 0.2cm
\noindent
{\it The matrix defined by $\cap_{ij}=\t\lb\chi\lp\pi(p_{i})\jj\pi(p_{j})\jj^{-1}\rp\rb$ has non vanishing determinant, where $p_{i}\in M_{n_{i}}(\kk)$ is a self-adjoint projection of minimal rank.}\\
\vskip 0.1 truecm

\subsection{First order condition}

In ordinary differential geometry of a compact manifold endowed with a spin structure, the Dirac operator is a first order differential operator. This can be easily formulated using commutation relations. Indeed, let $V$ be a compact manifold and $E$ and $F$ be smooth vector bundles over $V$. The space of sections $\Gamma(E)$ and $\Gamma(F)$ of the vector bundles turn out to be modules over $C^{\infty}(V)$ and one can easily characterize first order operators. They are the differential operators $D\, :\;\Gamma(E)\rightarrow\Gamma(F)$ that fulfill
$$
\lb\lb D,f\rb,g\rb=0,
$$   
for any couple of smooth functions $f,g\in C^{\infty}(V)$. This can also be extended to higher order differential operators: a differential operator $\dd$ is of order n if only if, for any $f\in C^{\infty}(V)$ $\lb\dd,f\rb$ is of order n-1.
\par
However, this notion cannot be extended in a direct manner in the noncommutative setting. In fact, we have to replace the modules by bimodules, and the functions in the double commutator by the left and right actions of the algebras. Since the latter commute, no matter whether the left or the right action appears first. 
\vskip 0.5truecm
\ul{{\bf First order condition}}
\vskip 0.2cm
\noindent
{\it $\lb\lb\dd,\pi(x)\rb,\jj\pi(y)\jj^{-1}\rb=0$ for any $x,y\in\aa$}.\\
\vskip 0.1truecm
We can rewrite the previous relation using bimodule notations. It becomes 
$$
\dd(x\Psi y)=x\dd(\psi y)+\dd(x\psi)y-x\dd(\psi)y,
$$
for any $\psi\in\hh$ and $x,y\in\aa$. This is the usual definition of a first order operator on a bimodule \cite{masson}.  

\subsection{$S^{0}$-reality}

The last four axioms define finite spectral triples. However, it turns out that the spectral triple of the standard model belongs to the smaller class of finite spectral triples that exclude Majorana particles. The operator $\jj$ can be considered as a charge conjugation operator and it is natural to ask whether there are or not Majorana fermions. Spectral triples without Majorana fermions, such as the spectral triple of the standard model, are called $S^{0}$-real spectral triples. They are defined by imposing a fifth axiom called.
\vskip 0.5truecm
\ul{{\bf $S^{0}$-reality}}
\vskip 0.2cm
\noindent
{\it There is a hermitian involution $\gamma$ that commutes with $\chi$, $\dd$ and $\pi(x)$ for any $x\in\aa$ and anticommutes with $\jj$.}\\
\vskip 0.1truecm
Since $\gamma$ is a hermitian involution, one has $\gamma^{*}=\gamma$ and $\gamma^{2}=1$, so that one can decompose $\hh$ into a direct sum of eigenspaces of $\gamma$. The eigenspace corresponding to eigenvalue +1 (resp. -1) is spanned by particles (resp. antiparticles). The anticommutation rule between $\gamma$ and $\jj$ teaches us that there is no particle that is its own antiparticle. In that case, $\hh$ is a direct sum,
$$
\hh=\hh_{L}^{P}\op\hh_{R}^{P}\op\hh_{L}^{A}\op\hh_{R}^{A},
$$ 
where P and A, refer to particle and antiparticle and L and R to left and right.
Within this decomposition, $\pi$ and $\chi$ are block diagonal,
\bbb
\pi&=&\mathrm{diag}\lp\pi_{L}^{P},\pi_{R}^{P},\pi_{L}^{A},\pi_{R}^{A}\rp,\n\\
\chi&=&\mathrm{diag}\lp -I_{n_{L}},I_{n_{R}},-I_{n_{L}},I_{n_{R}}\rp,\n
\eee
with $n_{L}=\mathrm{dim}(\hh_{L}^{P})=\mathrm{dim}(\hh_{L}^{A})$ and $n_{R}=\mathrm{dim}(\hh_{R}^{P})=\mathrm{dim}(\hh_{R}^{A})$. In a suitable basis, $\jj$ and $\dd$ are given by
$$
\dd=\pp{0&M&0&0\cr M^{*}&0&0&0\cr0&0&0&\ov{M}\cr0&0&\ov{M}^{*}&0\cr},\;
\jj=
\pp{0&0&I_{n_{L}}&0\cr0&0&0&I_{n_{R}}\cr I_{n_{L}}&0&0&0\cr0&I_{n_{R}}&0&0\cr}
\,C,
$$
where $C$ denotes the complex conjugation and $M\in M_{n_{L}\times n_{R}}(\cc)$ is a mass matrix. 
The reality and first order axioms are simply
\bbb
&\lb\pi_{L}^{P}(x),\ov{\pi}_{L}^{A}(y)\rb=0,&\n\\
&\lb\pi_{R}^{P}(x),\ov{\pi}_{R}^{A}(y)\rb=0,&\n\\
&\lp M\pi_{R}^{P}(x)-\pi_{L}^{P}(x)M\rp\ov{\pi}_{R}^{A}(y)=
\ov{\pi}_{L}^{A}(y)\lp M\pi_{R}^{P}(x)-\pi_{L}^{P}(x)M\rp,&\n
\eee
for any $x,y\in\aa$. Accordingly, particle and antiparticle spaces yield two bimodules, the latter being the opposite of the former in the sense that we exchange, up to complex conjugation, the left and right actions. Moreover, the Dirac operator, when restricted to the particle bimodule, is a first order operator, and it acts on antiparticle space as the complex conjugate. Therefore, the study of a $S^{0}$-real spectral triple can be reduced to study of the particle bimodule.

\section{General solution}

The axioms of noncommutative geometry turn out to be very simple in the finite case. Moreover, the algebra $\aa$ is a direct sum of matrix algebras and it becomes possible to give a general solution to these axioms. 

\subsection{The representation}

We first assume that the algebra is made of complex matrices with a complex representation, i.e. a representation linear over the complex numbers, and then, we extend our results to the real case. Let $\aa$ be a direct sum of complex matrix algebras,
$$
\aa=\mathop{\op}\limits_{i=1}^{N}M_{n_{i}}(\cc).
$$
To fix notations, we write elements $x,y$ of the algebra as N-uples: $x=(x_{i})_{1\leq i\leq N}$, with $x_{i}\in M_{n_{i}}(\cc)$. The only irreducible representation of $M_{n}(\cc)$ as a complex associative algebra is the fundamental one, so that $\pi$ can be reduced as
$$
\pi(x)=\mathop{\op}\limits_{i}x_{i}\ot I_{m_{i}},
$$
where, for $m\in\nn$, $I_{m}$ denotes the identity of $M_{m}(\cc)$. If $m=0$, we assume that $I_{m}$ is equal to zero and does not appear in the previous decomposition.
\par
The reality axiom provides us with two commuting representatioms of $\aa$, $x\mapsto\pi(x)$ and $y\mapsto\jj\pi(\ov{y})\jj^{-1}$, that can be simultaneously decomposed into irreducible ones. We end up with a {\it matrix of multiplicities}, $m\in M_{N}(\nn)$, and we have
\bbb
\pi(x)&=&\mathop{\op}\limits_{ij}
x_{i}\ot I_{m_{ij}}\ot I_{n_{j}},\n\\
\jj\pi(y)\jj^{-1}&=&\mathop{\op}\limits_{ij}
I_{n_{i}}\ot I_{m_{ij}}\ot \ov{y}_{j}.\n
\eee
To this reduction of $\pi$ is associated the corresponding decomposition of $\hh$ as a direct sum of mutually orthogonal subspaces,
$$
\hh=\mathop{\op}\limits_{ij}\hh_{ij},
$$
where $\hh_{ij}$ is isomorphic to $\cc^{n_{i}}\ot\cc^{m_{ij}}\ot\cc^{n_{j}}$. The action of $\pi(x)$ and $\jj\pi(y)\jj^{-1}$ on $\hh_{ij}$ are respectively $x_{i}\ot I_{m_{ij}}\ot I_{n_{j}}$ and $I_{n_{i}}\ot I_{m_{ij}}\ot \ov{y_{j}}$. Accordingly, $\hh_{ij}$ can be identified, as a bimodule, with $m_{ij}$ copies of $M_{n_{i}\times n_{j}}(\cc)$, where the left and right actions are simply left and right matrix multiplications.
\par
In the real case, the situation is a bit more tricky. The algebras $M_{n}(\rr)$ and $M_{n}(\hhh)$ have only one irreducible representation, the fundamental one, whereas $M_{n}(\cc)$ has two inequivalent irreducible representations, the fundamental one and its complex conjugate. However, all our formulae remain valid, provided the indices $i$ and $j$ label irreducible representations of $\aa$. In this case, $x_{i}$ denotes the image of $x\in\aa$ by the irreducible representation corresponding to $i$, and $\hh_{ij}$ is isomorphic to $\cc^{n_{i}}\ot\cc^{m_{ij}}\ot\cc^{n_{j}}$, where $n_{i}$ is the dimension of the irreducible representation $i$.
\par
To introduce the matrix of multiplicities $m$, we only use the bimodule structure of $\hh$. Thus, to any $(\aa,\bb)$-bimodule $\mm$, where $\mm$ is finite dimensional and $\aa$ and $\bb$ are direct sums of matrix algebras, we  associate a matrix of multiplicity $m$ by reducing simultaneously the left and right actions. The row index of $m$ labels the irreducible representations of $\aa$ whereas its column index labels those of $\bb$. Furthermore, it is possible to perform some algebraic operations on the matrices of multiplicities:
\begin{itemize}
\item
If $\mm$ is a $(\aa,\bb)$-bimodule with matrix $m$, the opposite $(\bb,\aa)$-bimodule corresponds to $m^{*}$.
\item
If $\mm$ and $\nnn$ are $(\aa,\bb)$-bimodules with matrices $m$ and $n$, the direct sum $\mm\op\nnn$ is a $(\aa,\bb)$-bimodule associated to the matrix $m+n$. \item
If $\mm$ is a $(\aa,\bb)$-bimodule with matrix $m$ and $\nnn$ a $(\bb,\ccc)$-bimodule with matrix $n$, the tensor product $\mm\ot_{\bb}\nnn$ is a $(\aa,\ccc)$-bimodule that corresponds to product $mn$. 
\end{itemize}
This is a very simple case of the theory of composition of correspondences, where $\aa$, $\bb$ and $\ccc$ are general Von Neumann algebras \cite{connes4,jones}. It applies in particular for a $S^{0}$-real spectral triple. The particle space is a bimodule associated to a matrix $e$, the antiparticle space is associated to $e^{*}$ and the total space corresponds to $m=e+e^*$, so that the matrix of multiplicities is symmetric. 
\par
{\bf Conclusion} A bimodule over a couple of semisimple involutive algebras is given, up to unitary equivalence, by a matrix of multiplicities whose entries are positive integers and whose row and column indices label irreducible representations of the corresponding algebras.

\subsection{The charge conjugation}

Given a $(\aa,\aa)$-bimodule $\hh$ with matrix $m$, it is natural to ask whether it corresponds to a real structure (that is, we can pass from one representation to the other by means of a charge conjugation $\jj$ as above) or not. Furthermore, if $\jj$ exists, it is interesting to determine its most general form.
\par
To answer these questions, let us first assume that it exists. The action of $\jj$ on $\psi\in\hh$ can always be written, in any basis, as $\jj\psi=K\ov{\psi}$, where $K$ is a matrix. $\jj$ is an antiunitary involution if only if $K\ov{K}=KK^{*}=1$. We must also have
$$
K\lp\mathop{\op}\limits_{i,j}x_{i}\ot I_{m_{ij}} \ot I_{n_{j}}\rp=
\lp\mathop{\op}\limits_{i,j}I_{n_{i}}\ot I_{m_{ij}} \ot x_{j}\rp K,
$$
for any $x\in\aa$. Accordingly,
$$
K\lp\mathop{\op}\limits_{j}\hh_{ij}\rp\subset\mathop{\op}\limits_{j}\hh_{ji},
$$ 
and since $K^{-1}=K^{*}$,
$$
K(\hh_{ij})=\hh_{ji}.
$$
Therefore, the dimensions of $\hh_{ij}$ and $\hh_{ji}$ are equal, or equivalently, the matrix of multiplicities $m$ is symmetric: $m_{ij}=m_{ji}$.
\par
Let us denote by $K_{ij}\, :\,\hh_{ij}\rightarrow\hh_{ji}$ the restriction of $K$. It must fulfill
\bbb
K_{ij}\lp x_{i}\ot I_{m_{ij}}\ot I_{n_{j}}\rp&=& 
\lp I_{n_{i}}\ot I_{m_{ij}}\ot x_{j}\rp K_{ij},\n\\
&K_{ij}K_{ij}^{*}=K_{ij}\ov{K_{ij}}=1.&\n
\eee
Consequently, if $e_{i}^{a}\ot\psi_{ij}^{p}\ot e_{j}^{b}$ is a basis of $\hh_{ij}=\cc^{n_{i}}\ot\cc^{m_{ij}}\ot\cc^{n_{j}}$, we have
\bbb
K_{ij}\lp e_{i}^{a}\ot\psi_{ij}^{p}\ot e_{j}^{b}\rp&=&
e_{j}^{b}\ot L_{ij}\psi_{ij}^{p}\ot e_{i}^{a}.\n\\
L_{ij}L_{ij}^{*}&=&1.\n
\eee
If $i\neq j$, the spaces $\hh_{ij}$ and $\hh_{ji}$ are distinct and it is always possible, by, a suitable choice of the orthonormal basis $(\psi_{ij}^{p})_{1\leq p\leq m_{ij}}$, to have $L_{ij}\psi_{ij}^{p}=\psi_{ji}^{p}$. When $i=j$, $L_{ii}$ is unitary, and we can choose an orthonormal basis such that $L_{ii}\psi_{ii}^{p}=e^{i\phi_{i}^{p}}\psi_{ii}^{p}$, with $\phi_{i}^{p}\in\rr$. However, since $\jj$ is antilinear,
\bbb
\jj\lp e^{i\phi_{i}^{p}/2}\,e_{i}^{a}\ot\psi_{ii}^{p}\ot e_{i}^{b}\rp&=&
e^{-i\phi_{i}^{p}/2}\,
K\lp \ov{e}_{i}^{a}\ot\ov{\psi}_{ii}^{p}\ot \ov{e}_{i}^{b}\rp,\n\\
&=& e^{-i\phi_{i}^{p}/2}\,
\ov{e}_{i}^{a}\ot L_{ii}\ov{\psi}_{ii}^{p}\ot\ov{e}_{i}^{b},\n\\
&=& e^{+i\phi_{i}^{p}/2}\,
\ov{e}_{i}^{a}\ot\ov{\psi}_{ii}^{p}\ot\ov{e}_{i}^{b},\n
\eee
so that the phases disappear if we make the replacement $\psi_{ij}^{p}\rightarrow e^{i\phi_{i}^{p}/2}\psi_{ij}^{p}$. Thus the action of $\jj$ is simply
$$
\jj\lp e_{i}^{a}\ot\psi_{ij}^{p}\ot e_{j}^{b}\rp=
\ov{e}_{j}^{b}\ot\ov{\psi}_{ji}^{p}\ot\ov{e}_{i}^{a},
$$
in all cases.
\par
Conversely, if the matrix of multiplicities $m$ is symmetric, it is possible to define $\jj$ as above and such a antilinear operator fulfills all conditions imposed to charge conjugation. 
\par
In the $S^{0}$-real case, we have already proved that $m=e+e^{*}$. In this case, we have two bimodules, one with matrix $e$ and the other with matrix $e^{*}$. If we define $\jj$ as the antilinear map that exchange the corresponding vectors of the two bimodules and $\gamma$ the hermitian operator that takes value 1 on the first bimodule and $-1$ on the second one, they anticommutes and the direct sum of the two bimodules corresponds, at the bimodule level (i.e. if we do not take care of $\chi$ and $\dd$), to a $S^{0}$-real structure.
\par
{\bf Conclusion}: a $(\aa,\aa)$-bimodule with matrix of multiplicities $m$ is endowed with a real structure if only if $m$ is symmetric. Moreover, if so, the charge conjugation is unique up to unitary equivalence, and it acts as above on tensor products. It corresponds to a $S^{0}$-real structure if only if $m$ can be written as $m=e+e^{*}$, with $e\in M_{N}(\nn)$, or equivalently, if all diagonal entries of $m$ are even.

\subsection{The chirality}

The chirality $\chi$ commutes with $\pi(\aa)$ and $\jj\pi(\aa)\jj^{-1}$, and is hermitian, so that it can be written as a block diagonal matrix,
$$
\chi=\mathop{\op}\limits_{ij}I_{n_{i}}\ot\chi_{ij}\ot I_{n_{j}}.
$$
Since $\chi=\sum_{p}\pi(x^{p})\jj\pi(y^{p})\jj^{-1}$, with $x^{p},y^{p}\in\aa$,  $\chi_{ij}$ is scalar matrix. Furthermore $\chi_{ij}=\pm1$ as follows from $\chi^{2}=1$. We also have $\lb\chi,\jj\rb=0$ so that $\chi_{ij}=\chi_{ji}$.
\par
All the information contained in the matrices $\chi_{ij}$ and $m_{ij}$ can be cast in a symmetric matrix $\mu\in M_{N}(\zz)$ defined by 
$$
\mu_{ij}=\chi_{ij}m_{ij}.
$$ 
We recover $\chi_{ij}$ and $m_{ij}$ as the sign and the modulus of $\mu_{ij}$. Accordingly,
$$
\chi=\mathop{\op}\limits_{ij}\mathrm{sign}(\mu_{ij})I_{n_{i}}\ot I_{|\mu_{ij}|}\ot I_{n_{j}}.
$$
\par
To proceed further, let us assume that we are dealing with the complex case. Thus all projections involved in Poincar\'e duality are of trace 1, and the matrix of the intersection form $\cap_{ij}$ is given by
\bbb
\cap_{ij}&=&\t\lb\chi\pi(p_{i})\jj\pi(p_{j})\jj^{-1}\rb,\n\\
&=&\mathrm{sign}(\mu_{ij})\t\lb p_{i}\ot I_{|\mu_{ij}|}\ot \ov{p}_{j}\rb,\n\\
&=&\mu_{ij}.\n
\eee
According to Poincar\'e duality, the matrix $\mu$ must be non degenerate.
\par
Conversely, if we start with a symmetric and non degenerate matrix $\mu\in M_{N}(\zz)$ (from now on we refer to it as the matrix of multiplicities instead of $m$), we recover the bimodule structure out of $|\mu_{ij}|$ and the chirality $\chi$ is obtained as
$$
\chi=\mathop{\op}\limits_{ij}\mathrm{sign}(\mu_{ij})I_{n_{i}}\ot I_{|\mu_{ij}|}\ot I_{n_{j}}.
$$
It fulfills all the requirements imposed by Poincar\'e duality and by the orientability axiom. For instance, let us check that there are some elements, $x^{p}$ and $y^{p}$, of $\aa$ such that $\chi=\sum_{p}\pi(x^{p})\jj\pi(y^{p})\jj^{-1}$. To proceed, let us expand the matrix $\chi_{ij}=\mathrm{sign}(\mu_{ij})$ over rank one matrices as
$$
\chi_{ij}=\mathop{\sum}\limits_{p}x_{i}^{p}\ov{y}_{j}^{p}.
$$         
Thus we have $\chi=\sum_{p}\pi(x^{p})\jj\pi(y^{p})\jj^{-1}$, with $x^{p}=(x_{i}^{p}1_{i})_{1\leq i\leq N}$ and $y^{p}=(y_{i}^{p}1_{i})_{1\leq i\leq N}$.
\par
Accordingly, in the complex case, the bimodule structure and the chirality of a finite spectral triple are given, up to unitary equivalence, by a symmetric and non degenerate matrix $\mu$ with entries in $\zz$, which is the matrix of the intersection form. Furthermore, it corresponds to a $S^{0}$-real structure if only if the intersection form is an even quadratic form. In the real case, it is still given by a symmetric matrix with entries in $\zz$ whose indices label irreducibles representations of $\aa$. However, it is no longer the intersection form and the signs of its entries satisfy additional requirements that are summarized, together with its relation with the intersection form, in table 1.        

\begin{center}
\begin{tabular}{|c|c|c|c|} 
\hline
&$\rr$&$\cc$&$\hhh$\\
\hline
&&&\\
&&$\mu_{ij\alpha}\mu_{ij\bar{\alpha}}\geq 0$&\\
$\rr$&&&\\
&$\cap_{ij}=\mu_{ij}$
&$\cap_{ij}=\mathop{\sum}\limits_{\alpha}\mu_{ij\alpha}$
&$\cap_{ij}=2\mu_{ij}$\\
&&&\\
\hline
&&&\\
&$\mu_{i\alpha j}\mu_{i\bar{\alpha}j}\geq 0$
&$\mu_{i\alpha j\beta}\mu_{i\bar{\alpha} j\bar{\beta}}\geq 0$
&$\mu_{i\alpha j}\mu_{i\bar{\alpha} j}\geq 0$\\
$\cc$&&&\\
&$\cap_{ij}=\mathop{\sum}\limits_{\alpha}\mu_{i\alpha j}$
&$\cap_{ij}=\mathop{\sum}\limits_{\alpha\beta}\mu_{i\alpha j\beta}$
&$\cap_{ij}=2\mathop{\sum}\limits_{\alpha}\mu_{i\alpha j}$\\
&&&\\
\hline
&&&\\
&&$\mu_{ij\alpha}\mu_{ij\bar{\alpha}}\geq 0$&\\
$\hhh$&&&\\
&$\cap_{ij}=2\mu_{ij}$
&$\cap_{ij}=2\mathop{\sum}\limits_{\alpha}\mu_{ij\alpha}$
&$\cap_{ij}=4\mu_{ij}$\\
&&&\\
\hline
\end{tabular}
\end{center}
\vskip 0.5truecm
\centerline{\ul{Table 1}}
\vskip 0.5truecm
In this table, the indices $i$ and $j$ label, for matrices with entries in a given field, the simple algebras that appear in the decomposition of $\aa$. In the complex case, the indices $\alpha$ and $\beta$ distinguish a representation from its complex conjugate.
\par
{\bf Conclusion}: The bimodule structure and the the chirality of a finite spectral triple are given, up to unitary equivalence, by a symmetric and non degenerate matrix of multiplicities. The moduli of the entries of the matrix of multiplicities determine the bimodule structure and their signs yield the chirality. 

\subsection{The Dirac operator}

As we have already pointed out, a first order operator $D$ on a $(\aa,\bb)$-bimodule $\mm$ is an operator that satisfies
$$
D(x\psi y)= xD(\psi y)+D(x\psi)y-xD(\psi)y,
$$
for any $\psi\in\mm$, $x\in\aa$ and $y\in\bb$. If we denote by $L_{L}(\mm)$ (respectively $L_{R}(\mm)$, $L_{1}(\mm)$) the space of left $\aa$-linear operators (respectively right $\bb$-linear operators, first order operators), we have
$$
L_{L}(\mm)+L_{R}(\mm)\subset L_{1}(\mm).
$$
If $\mm$ is finite dimensional and if the algebras are direct sum of matrix algebras, the converse also holds 
$$
L_{L}(\mm)+L_{R}(\mm)=L_{1}(\mm).
$$
To prove it, we decompose the representations associated with the left and right actions into irreducible ones and determine the corresponding matrix elements of $D$.
\par
In particular, this applies to the Dirac operator $\dd$ of a finite spectral triple that can be written as
$$
\dd=\dd_{L}+\dd_{R}
$$
with
$$
\lb\dd_{L},\pi(\aa)\rb=\lb\dd_{R},\jj\pi(\aa)\jj^{-1}\rb=0.
$$
Moreover, since $\dd$ anticommutes with $\chi$, so do $\dd_{L}$ and $\dd_{R}$. Accordingly, the previous decomposition is unique. Indeed, if we have, with obvious notations, 
$$
\dd=\dd_{L}^{1}+\dd^{1}_{R}=\dd^{2}_{L}+\dd^{2}_{R},
$$
then
$$
\dd^{1}_{L}-\dd^{2}_{L}=\dd^{2}_{R}-\dd^{1}_{R}
$$
commutes with both $\pi(\aa)$ and $\jj\pi(\aa)\jj^{-1}$. Thus, it commutes with $\chi$, and since it also anticommutes with $\chi$, it vanishes and the previous decomposition is unique. Furthermore, $\dd$ commutes with $\jj$, so that, using the unicity, we have
$$
\dd_{L}=\jj\dd_{R}\jj^{-1}.
$$
\par
Consequently, the Dirac operator of a finite spectral triple is given by
$$
\dd=\Delta+\jj\Delta\jj^{-1},
$$
where $\Delta$ is a hermitian operator on $\hh$, which commutes with $\jj\pi(\aa)\jj^{-1}$ and anticommutes with $\chi$. For a $S^{0}$ real spectral triple, it is sufficient to take into account the particle bimodule. The Dirac operator can be written as a direct sum,
$$
\dd=\dd_{P}\op\dd_{A},
$$
where $\dd_{P}$ is a first order operator of the particle bimodule. Using the unicity, we have
$$
\dd_{A}=\ov{\dd}_{P}.
$$
The decomposition of $\dd$ remains valid for the bimodule structure of the particle space, 
$$
\dd_{P}=\Delta_{P}+\ov{\Delta}_{A},
$$
with
$$
\lb\Delta_{A},\pi_{A}(\aa)\rb=\lb\Delta_{P},\pi_{P}(\aa)\rb=0.
$$
\par 
Let us end this section by three remarks that are useful when this construction  is applied to model building in gauge theory. First of all, we can give an explicit expression of $\Delta$ in terms of the Dirac operator. Indeed,
$$
\dd=-\int_{G}dg\,\pi(g)\lb\dd,\pi(g^{*})\rb+\int_{G}dg\,\pi(g)\dd\pi(g^{*}),
$$
where $dg$ is the Haar measure on the group $G$ of unitary elements of $\aa$. Due to the translational invariance of the Haar measure,
$$
\lb\int_{G}dg\,\pi(g)\dd\pi(g^{*}),\pi(x)\rb=0,
$$ 
for any $x\in\aa$. Futhermore, it follows from the reality axiom and the first order condition that
$$
\lb\int_{G}dg\,\pi(g)\lb\dd,\pi(g^{*})\rb,\jj\pi(x)\jj\rb=0,
$$ 
for any $x\in\aa$. Thus, it is a decomposition of the Dirac operator into left and right module homomorphisms that both anticommute with $\chi$. Using the unicity, we have
$$
\Delta=-\int_{G}\,dg\,\pi(g)\lb\dd,\pi(g^{*})\rb.
$$
This expression of $\Delta$ is very useful since it means that $\Delta$ is a 1-form.
\par
Secondly, The operator $\Delta$ satisfies very simple commutation and anticommutation relations and we can give its explicit form in terms of complex matrices. Let
$$
\hh=\mathop{\op}\limits_{ij}\hh_{ij}
$$  
be the decomposition of the Hilbert where $i$ and $j$ label irreducible representations of $\aa$, and $P_{ij}\;\hh\rightarrow\hh_{ij}$ be the orthogonal projection onto $\hh_{ij}$. We define the matrix elements of $\Delta$ by
$$
\Delta_{ij}^{kl}=P_{kl}\,\Delta\,P_{ij}^{*}.
$$ 
The latter are given by
$$
\Delta_{ij}^{kl}=\delta_{jl}\,M_{ik,j}\ot I_{n_{j}},
$$
where $M_{ik,j}\in M_{\mu_{ij}n_{i}\times\mu_{kj}n_{k}}(\cc)$ is such that $M_{ik,j}=M_{ki,j}^{*}$ if $\mu_{ij}\mu_{kj}<0$, and zero otherwise. This gives us a complete description of the operator $\Delta$ parametrized by complex matrices, and we recover the Dirac operator as $\dd=\Delta+\jj\Delta\jj^{-1}$. 
\par
Thirdly, notice that most of the time, the Dirac operator $\dd$ only appears in noncommutative geometry via the commutation relation $\lb\dd,\pi(x)\rb$, $x\in\aa$. Since $\dd=\Delta+\jj\Delta\jj^{-1}$ and $\lb\jj\Delta\jj^{-1},\pi(x)\rb=0$ for any $x\in\aa$, one can replace $\dd$ by $\Delta$ in any commutation relation with an element of the algebra. This simplifies, for finite spectral triples, many computations involving the Dirac operator, such as, for instance, the computation of the metric on the space of pure states of $\aa$ and its pertubation by inner fluctuations \cite{connes5}. 
\par
Finally, we use this decomposition to compute the 1-forms. We recall that the space of 1-forms $\Omega_{\dd}^{1}(\aa)$ is defined by
$$
\Omega_{\dd}^{1}(\aa)=\la\mathop{\sum}\limits_{p}\pi(x^{p})\lb\dd,\pi(y^{p})\rb\;\mathrm{with}\;x^{p},y^{p}\in\aa\ra.
$$ 
In particular,
$$
\Delta=-\int dg\;\pi(g)\lb\dd,\pi(g^{*})\rb,
$$
is a 1-form. Moreover, if $\omega\in\Omega_{\dd}^{1}(\aa)$ and $x,y\in\aa$, we have  $\pi(x)\omega\pi(y)\in\Omega_{\dd}^{1}(\aa)$ and   $\Omega_{\dd}^{1}(\aa)$ is a bimodule and thus contains the bimodule generated by $\Delta$. Conversely, if we express any element  $\omega\in\Omega_{\dd}^{1}(\aa)$ using commutators with $\dd$ and then, replace $\dd$ by $\Delta$, we see that $\Omega_{\dd}^{1}(\aa)$ is included in the bimodule generated by $\Delta$. Accordingly, the space of 1-forms  $\Omega_{\dd}^{1}(\aa)$ is the bimodule generated by $\Delta$, and this allows us, using the expression of $\Delta$ in terms of complex matrices, to determine  $\Omega_{\dd}^{1}(\aa)$ in section 5. 
\par
{\bf Conclusion}: A finite spectral triple is given, up to unitary equivalence by a symmetric and non degenerate matrix of multiplicities and a hermitian operator $\Delta$ commuting with the right action and anticommuting with the chirality.

\section{Classification and diagrams}

In the following sections, we develop a diagrammatic approach to finite spectral triples. We first associate to any finite spectral triple a diagram and then, we  classify and construct all such triples within this approach.

\subsection{Diagram associated with a spectral triple}

Let $(\aa,\hh,\dd)$ be a finite spectral triple whose matrix of multiplicities is $\mu$. To each couple $(i,j)$ of irreducible representations of $\aa$ such that $\mu_{ij}\neq 0$, we associate a vertex of type $\om$ if $\mu_{ij}<0$ and a vertex of type $\op$ if $\mu_{ij}>0$. We relate the vertices $(i,j)$ and $(k,l)$ if only if the corresponding matrix element of the Dirac operator,
$$
\dd_{ij}^{kl}=P_{ij}\,\dd\,P_{kl}^{*},
$$
does not vanish. According to the axioms, the diagram has the following properties.
\begin{itemize}
\item
$(i,j)$ is related to $(k,l)$ if only if $(k,l)$ is related to (i,j) as follows from $\dd=\dd^{*}$. To simplify, we draw only one edge between two vertices.
\item
Since $\dd$ and $\jj$ commute, the diagram is symmetric with respect to its first diagonal.
\item
We can only relate vertices of different types, as follows from the anticommutation relation of $\chi$ and $\dd$. 
\item
As a consequence of the first order condition, all edges are either vertical or               horizontal. Vertical edges correspond to matrix elements of $\Delta$ and the horizontal ones to those of $\jj\Delta\jj^{-1}$.
\end{itemize}
\par
For instance, let us build the diagram pertaining to the standard model. The algebra $\aa$ is 
$$
\aa=\hhh\op\cc\op M_{3}(\cc).
$$
Since it is $S^{0}$-real, we use the notations introduced in section 2.5. The  particle Hilbert spaces $\hh_{L}^{P}$ and $\hh_{R}^{P}$ are spanned, for $N_{f}=3$ families of fermions, by  
$$
\pp{u\cr d}_{L},\,
\pp{c\cr s}_{L},\,
\pp{t\cr b}_{L},\,
\pp{\nu_{e}\cr e}_{L},\,
\pp{\nu_{\mu}\cr \mu}_{L},\,
\pp{\nu_{\tau}\cr \tau}_{L},
$$
and
$$
(u)_{R},\, (d)_{R},\, (c)_{R},\, (s)_{R},\, (t)_{R},
\, (b)_{R},\, (e)_{R},\, (\mu)_{R},\,(\tau)_{R},  
$$
where we have omitted the color index for quarks. The corresponding antiparticles form a basis of the antiparticle spaces $\hh_{L}^{A}$ and $\hh_{R}^{A}$. Within these bases, the representation is given by
\bbb
\pi_{L}^{P}(a)&=&\mathrm{diag}\lp a\ot I_{3N_{f}},\,a\ot I_{N_{f}}\rp\nonumber,\\
\pi_{R}^{P}(b)&=&\mathrm{diag}\lp b\,I_{3N_{f}},\,\ov{b}\,I_{3N_{f}},\,\ov{b}\,I_{N_{f}}\rp\nonumber,\\
\pi_{L}^{A}(b,c)&=&\mathrm{diag}\lp I_{2N_{f}
}\ot c,\,b\,I_{2N_{f}}\rp\nonumber,\\
\pi_{R}^{A}(b,c)&=&\mathrm{diag}\lp
I_{2N_{f}}\ot c,\,b\,I_{N_{f}}\rp\nonumber,
\eee
where $(a,b,c)\in\hhh\op\cc\op M_{3}(\cc)$. The mass matrix $M$ is $$
M=\pp{\pp{M_{u}\ot I_{3}&0\cr 0&M_{d}\ot I_{3}}&0\cr
0&\pp{0\cr M_{e}}},
$$
with 
\bbb
M_{u}&=&\mathrm{diag}\lp m_{u},m_{c},m_{t}\rp,\n\\
M_{d}&=&V_{CKM}\,\mathrm{diag}\lp m_{d},m_{s},m_{b}\rp,\n\\
M_{e}&=&\mathrm{diag}\lp m_{e},m_{\mu},m_{\tau}\rp,\n
\eee
where $m_{p}$ stands for the mass of particle $p$ and $V_{CKM}$ is the Cabibbo-Kobayashi-Maskawa mixing matrix.
\par
Accordingly, the matrix of multiplicities is, in the basis $(\cc,\hhh,\ov{\cc},M_{3}(\cc))$,
$$
\mu=N_{f}\,\pp{0&0&1&1\cr 0&0&-1&-1\cr 1&-1&0&1\cr 1&-1&1&0\cr},
$$
and the corresponding diagram is given by figure 1.
\begin{figure}
\centering
\epsfig{file={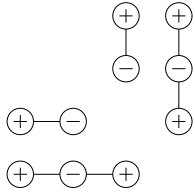},width=2.5cm}
\caption{Diagram of the standard model}
\end{figure}
\par

For particles, vertices and edges are above the first diagonal. The three vertices $\op$ correspond to the singlets of isospin $u_{R}$, $d_{R}$ and $e_{R}$ and the two vertices $\om$ are associated with the leptonic and quarkonic isodoublets. The edges between them correpond to the matrix elements $M_{u}$, $M_{d}$ and $M_{e}$. Since it is $S^{0}$-real, there is a complete symmetry between particles and antiparticles so that the same holds for the antiparticles. In fact, for a general $S^{0}$-real spectral triple, the matrix of multiplicities can always be written as
$$
\mu=\epsilon+\epsilon^{*}.
$$ 
The Dirac operator is a first order operator of the particle bimodule and the complex conjugate on the antiparticle bimodule. Thus, the corresponding diagram is constructed out of the vertices of the matrix $\epsilon$ and then symmetrized with respect to the first diagonal.
\par
The diagram of the standard model is not connected and has four connected subdiagrams that are intimately tied up with the "noncommutative coupling constants" $x$, $y$, $\tilde{x}$ and $\tilde{y}$, introduced in \cite{carminati}. In general, the latter is a positive definite hermitian matrix $Z$ that commutes with $\pi(\aa)$, $\jj\pi(\aa)\jj^{-1}$ and $\dd$, so that it yields a scalar product on the differential algebra. The first two commutation relations allow us to write $Z$ as a block diagonal matrix,
$$
Z=\mathop{\op}\limits_{ij} I_{n_{i}}\ot Z_{ij}\ot I_{n_{j}}.
$$
Accordingly, to each vertex $(i,j)$ is associated a positive definite hermitian matrix $Z_{ij}$. The last commutation relation, expressed using matrix elements as 
$$
Z_{ij}\,\dd_{ij}^{kl}=\dd_{ij}^{kl}\, Z_{kl},
$$
relates the matrix elements $Z_{ij}$ associated with vertices belonging to the same connected component.
\par
The diagrams provides us with a natural classification of finite spectral triples. As we shall see in the sections devoted to applications, the major feature of the underlying Yang-Mills-Higgs model can be seen on the diagram. However, a commutative spectral triple, with the algebra $\cc^{N}$ and a noncommutative one, whose algebra is a direct sum of $N$ simple matrix algebras, can yield the same diagram. From the point of view of noncommutative geometry, this is not surprising since two such algebras are Morita equivalent and thus completely indistinguishable in Hoschild homology and $K$-theory. Besides, using the general construction, based on a connection $\nabla$ on a finite projective module $\ee$, detailed in \cite{connes5}, we can pass from a commutative spectral triple to a noncommutative one which are both associated with the same diagram.   
 
\subsection{Spectral triple associated to a diagram}

Suppose we are given an algebra $\aa$ and a symmetric matrix of multiplicities $\mu$ whose indices label irreducible representations of $\aa$. We require that it fulfills the constraints listed in table 1  and that its intersection form $\cap_{ij}$ be non degenerate. We represent it on the plane as the vertices of a diagram as we did in the previous section. Then, we relate the vertices of different types by vertical and horizontal lines such that the resulting diagram be symmetric with respect to its first diagonal.
\par
To each vertical link between $(i,k)$ and $(j,k)$, with $i>j$, we associate a matrix $M_{ij,k}\in M_{\mu_{ik}n_{i}\times \mu_{jk}n_{j}}(\cc)$. If $j>i$, we set $M_{ji,k}=M_{ij,k}^{*}$ and $M_{ij,k}=0$ in all other cases. The matrix elements of the operator $\Delta$ are obtained as
$$
\Delta_{ik}^{jl}=\delta_{kl}\, M_{ij,k}\ot I_{n_{k}}.
$$
In particular, the diagonal elements $\Delta_{ij}^{ij}$ always vanish since we cannot a relate a vertice to itself. We symmetrize this expression with respect to $\jj$ to get the complete Dirac operator
$$
\dd=\Delta+\jj\Delta\jj^{-1}.
$$  
To obtain an explicit form of $\dd$ in terms of complex matrices, we introduce a basis $(E_{ij}^{p})_{1\leq p\leq n_{i}n_{j}}$, for example the canonical one, and we write $M_{ij,k}$ as
$$
M_{ij,k}=\mathop{\sum}\limits_{p} E_{ij}^{p}\ot M_{ij,k}^{p}.
$$
Thus, the matrix elements of the Dirac operator are
$$
\dd_{ik}^{jl}=\delta_{kl}\;\mathop{\sum}\limits_{p} 
E_{ij}^{p}\ot M_{ij,k}^{p}\ot I_{n_{k}}+\delta_{ij}\;\mathop{\sum}\limits_{q} 
I_{n_{i}}\ot\ov{M}_{kl,i}^{p}\ot\ov{E}_{kl}^{q}.
$$
The representation, the Hilbert space and the operators $\jj$ and $\chi$ are constructed out of the matrix of multiplicities by means of the formulae we gave throughout the sections 3.1-3.3.

\section{Applications to particle models}

Finite spectral triples are very simple examples of noncommutative geometries corresponding to finite spaces. The latter can be considered as the internal spaces involved in Yang-Mills-Higgs theories. More precisely, if we tensorize a finite spectral triple with the one associated to an ordinary manifold of dimension n, we obtain a noncommutative geometry of dimension n  that gives rise to a Yang-Mills theory with spontaneous symmetry breaking. This section is devoted to this kind of applications. We first recall some general aspects about gauge theory in noncommutative geometry and then apply our previous classification to investigate some features of the model we obtain.

\subsection{General aspects}

It follows from general principles of noncommutative geometry \cite{connes5}, that the symmetries of a spectral triple $(\aa,\hh,\dd)$ are described by $*$-automorphisms of the algebra $\aa$. Among them, there are, in the noncommutative case, inner automorphisms. The latter are defined, as $x\in\aa\mapsto gxg^{-1}\in\aa$, for any element $g$ of the group $G$ of unitary elements of $\aa$. If the algebra is finite dimensional, the inner automorphisms are the only automorphisms of $\aa$ up to a discrete subgroup. Thus, they are the only possible continuous symmetries of a finite noncommutative space. To describe a symmetry of the corresponding space, the group $G$ is represented on the Hilbert space $\hh$ via the map $g\mapsto\pi(g)\jj\pi(g)\jj^{-1}$. 
The inner automorphisms are in complete analogy with gauge transformations
\par
In the framework of noncommutative geometry, $\hh$ is the fermionic Hilbert space and $\dd$ the Dirac operator. Accordingly, the free fermionic action is defined as
$$
S_{F}\lb\Psi\rb=\langle\Psi,\dd\Psi\rangle,
$$
for any $\Psi\in\hh$. Under a gauge transformation,
$$
\Psi\longrightarrow\pi(g)\jj\pi(g)\jj^{-1}\Psi,
$$
the fermionic action becomes
$$
\langle\Psi,\lp\dd+\pi(g^{*})\lb\dd,\pi(g)\rb
+\jj\pi(g^{*})\lb\dd,\pi(g)\rb\jj^{-1}\rp\Psi\rangle,
$$
where we have used the axioms to rewrite 
$$
\pi(g^{*})\jj\pi(g^{*})\jj^{-1}\;\dd\;\pi(g)\jj\pi(g)\jj^{-1}
$$ as 
$$
\dd+\pi(g^{*})\lb\dd,\pi(g)\rb+\jj\pi(g^{*})\lb\dd,\pi(g)\rb\jj^{-1}.
$$
To ensure gauge invariance, we introduce, as usual in field theory, a gauge field $A$. The latter are hermitian 1-forms and can be written as
$$
A=\mathop{\sum}\limits_{p}\pi(x^{p})\lb\dd,\pi(y^{p})\rb,
$$
with $x^{p},y^{p}\in\aa$. The transformation law of $A$ is
$$
A\rightarrow\pi(g)A\pi(g^{*})+\pi(g)\lb\dd,\pi(g^{*})\rb,
$$
so that the interacting fermionic action
$$
S_{F}\lb\Psi,A\rb=\langle\Psi,\lp D+A+\jj A\jj^{-1}\rp\Psi\rangle
$$
is gauge invariant. Starting with the spectral action principle, "the bosonic action only depends on the spectrum of $\dd+A+\jj A\jj^{-1}$", it is possible to write a bosonic action for the field $A$. With a suitable spectral triple, it yields, as first terms of heat kernel expansion, the bosonic sector of the standard model coupled to gravity \cite{connescham}.
\par
To construct a Yang-Mills-Higgs theory, we consider a noncommutative geometry $(\aa_{t},\hh_{t},\dd_{t})$ obtained as a product of the ordinary geometry of space time by a finite noncommutative one, described by a finite spectral triple $(\aa,\hh,\dd)$, that corresponds to the internal degrees of freedom of the fields. $(\aa_{t},\hh_{t},\dd_{t})$ is a tensor product of spectral triples,
\bbb
\aa_{t}&=&C^{\infty}(V)\ot\aa,\n\\
\hh_{t}&=&L^{2}(V,S)\ot\hh,\n\\
\dd_{t}&=&i\gamma^{\mu}\lp\partial_{\mu}+\omega_{\mu}\rp\ot I+\gamma^{5}\ot\dd,\n
\eee
where $\omega_{\mu}$ is the spin connection introduced to take into account the coupling of fermionic fields with gravity. Therefore, the gauge fields $A_{t}$ are given by
$$
A_{t}=i\gamma^{\mu}\lp A_{\mu}+\jj A_{\mu}\jj^{-1}\rp
+\gamma^{5}\ot\lp H+\jj H\jj^{-1}\rp,
$$
where $A^{\mu}$ is a vector field with values in the Lie algebra $\gg$ of the group of unitary elements of $\aa$, and $H$ is a scalar field that is also a hermitian 1-form with respect to the finite geometry. The corresponding fluctuations of the metric \cite{connes5} are
\bbb
&\dd_{t}+A_{t}+\jj_{t}A_{t}\jj_{t}=&\n\\
&i\gamma^{\mu}\lp \partial_{\mu}+\omega_{\mu}+A_{\mu}+\jj A_{\mu}\jj^{-1}\rp
+\gamma^{5}\ot\lp \dd+H+\jj H\jj^{-1}\rp.&\n
\eee
\par
Using the decomposition $\dd=\Delta+\jj\Delta\jj^{-1}$, the scalar part can be written as 
$$
\gamma^{5}\lp\Phi+\jj\Phi\jj^{-1}\rp,
$$
with $\Phi=\Delta+H$. Moreover, since $\Delta$ is a 1-form, $\Phi$ is a gauge field and it transforms as
$$
\Phi\longrightarrow\pi(g)\Phi\pi(g^{*}).
$$
The interaction of fermions with scalars can be written using $\Phi$ as
$$
\langle\Psi,\lp\Phi+\jj\Phi\jj^{-1}\rp\Psi\rangle.
$$
Thus, $\Phi$ is a matrix valued scalar field that contains all Higgs fields, and the previous interaction term corresponds to Yukawa couplings. 
\par
Finally, following the spectral action principle, the Higgs potential $V(\Phi)$ only depends on the spectrum of $\Phi+\jj\Phi\jj^{-1}$ and thus can be expanded
as
$$
V(\Phi)=\mathop{\sum}\limits_{n}\frac{a_{2n}}{(2n)!}
\t\lp\Phi+\jj\Phi\jj^{-1}\rp^{2n},
$$
since $\lp\Phi+\jj\Phi\jj^{-1}\rp^{2n+1}$ is traceless. In dimension four, we restrict our study to polynomial potential of degree $\leq 4$, so that
$$
V(\Phi)=a_{0}+
\frac{a_{2}}{2!}\t\lp\Phi+\jj\Phi\jj^{-1}\rp^{2}+
\frac{a_{4}}{4!}\t\lp\Phi+\jj\Phi\jj^{-1}\rp^{4}
$$
where $a_{0}$, $a_{2}$ and $a_{4}$  are real numbers related to other parameters of the theory using the spectral action principle applied to the full Dirac operator.

\subsection{The gauge group}

The gauge transformations correspond to the inner automorphisms of the finite dimensional algebra $\aa$. Accordingly, the gauge group is the group $G$ of unitary elements of $\aa$. Since $\aa$ is a direct sum of matrix algebras over the fields $\rr$, $\cc$ and $\hhh$, the gauge group is a direct product of the simple Lie groups $SO(n)$, $U(n)$ and $SP(n)$. As a consequence, noncommutative geometric models can have all classical semi-simple compact Lie group as a gauge group, except those corresponding to the five exceptional Lie algebras \cite{iochumschuker}.
\par
However, the gauge group only corresponds to $G$ up to some $U(1)$ factors. Indeed, the elements of $G$ appear in gauge transformations as $\pi(g)\jj\pi(g)\jj^{-1}$ so that we have to divide $G$ by the kernel of the representation  $g\mapsto\pi(g)\jj\pi(g)\jj^{-1}$ to obtain the gauge group. The latter only contains $U(1)$ factors, that corespond to the $U(1)$ whose representations only appear on diagonal vertices in the diagram associated to $(\aa,\hh,\dd)$. These $U(1)$ gauge fields also disappear from the action because the action only depends on $\pi(A^{\mu})+\jj\pi(A^{\mu})\jj^{-1}$ and they are the solutions of the equation $\pi(A^{\mu})+\jj\pi(A^{\mu})\jj^{-1}=0$.
\par
For example, let us consider a model with fermions in the adjoint representation. It is built with $\aa=M_{n}(\cc)$ and $\hh=M_{n}(\cc)$. The representation of $\aa$ on $\hh$ is the left multiplication of matrices and $\jj$ is given by the adjoint. Obviously, it corresponds to a diagram with a single vertex so that its gauge group is $SU(n)$ instead of $U(n)=SU(n)\times U(1)$. Moreover, from vertex uniqueness, we see immediately that it can only afford fermions with same chirality and there is no possible link, thus no Dirac operator and no Higgs field. Corresponding to the simplest diagram, this model can be considered as the minimal one. It can be extended to semi-simple algebras, with bimodule structure given by left and right multiplication. It yields a diagonal matrix of multiplicities and a diagonal diagram. Accordingly, all $U(1)$ disappear but there is still no possible links, thus no Higgs fields and no spontaneous symmetry breaking. For the standard model (see Figure 1), the diagram has no diagonal elements. Therefore, the unwanted $U(1)$ in the group of unitary elements of $\hhh\op\cc\op M_{3}(\cc)$ does not disappear and has to be eliminated by hand. 
\par
In more sophisticated models, we often have to remove by hand some $U(1)$ factors for physical reasons such as anomaly cancellation (of course, it may also be necessary to remove non-abelian Lie subgroups but this will change too much the noncommutative geometric setting of the model). This is achieved by means of a unimodularity condition, that can be defined as a set of linear conditions imposed at the Lie algebra level in order to eliminate the unwanted $U(1)$ factors. For instance, if we want our model to be gauge anomaly free, we impose that the generators of the Lie algebra of the gauge group, satisfy the cubic relation 
$$
\t\lp T^{a}\la T^{b},T^{c}\ra\rp=0.
$$
Since the representation we deal with are rather simple (tensor products of fundamental ones or their complex conjugates), the gauge anomaly cancellation can be expressed as constraints on the matrix of multiplicities of the model. For instance, let us consider a $S^{0}$-real spectral triple with the complex algebra 
$$
\aa=\mathop{\op}\limits_{i}M_{n_{i}}(\cc),
$$
and a complex representation. Taking into account the occurence of left and right-handed fermions, the condition of anomaly is simply
$$
\tp\lb\chi\lp\pi(x)+\jj\pi(x)\jj^{-1}\rp^{3}\rb=0\;\mathrm{for}\;x\in\gg,
$$
where $\gg$ is the Lie algebra of the gauge group and $\tp$ denotes the trace restricted to particle space. To express it in terms of the matrix of multiplicities, we write any $x_i\in u(n_{i})$ (Lie algebra of $U(n_{i})$) as $x_i=\lambda_{i}+u_{i}$, with $\lambda_{i}\in su(n_{i})$ and $u_{i}\in u(1)$. Since we only want to remove some $U(1)$ factors, we require that this condition be satisfied at least for all non abelian factors. It yields three different type of constraints on the matrix of multiplicities $\epsilon$.
\bbb
&\mathop{\sum}\limits_{j}
n_{j}\lp\epsilon_{ij}-\epsilon_{ji}\rp=0
\;\;\mathrm{if}\;n_{i}\geq3,&\n\\
&\mathop{\sum}\limits_{j}
n_{j}\lp\epsilon_{ij}-\epsilon_{ji}\rp\lp u_{i}-u_{j}\rp=0
\;\;\mathrm{if}\;n_{i}\geq2,&\n\\
&\mathop{\sum}\limits_{ij}
n_{i}n_{j}\lp\epsilon_{ij}-\epsilon_{ji}\rp\lp u_{i}-u_{j}\rp^{3}=0
.&\n
\eee
The first equations are rigid constraints that must be satisfied by the matrix of multiplicities. Ths second ones are linear equations for the $u_{i}$ 's so that they may be taken as unimodularity conditions. The last one is a cubic relation on the "hypercharges", as in the standard model. It is a well known fact that for the standard model, anomaly cancellation is equivalent to the unimodularity condition \cite{garcia}. In other words, if the second relations are satisfied, the third one also is. Obviously, this is not a general feature of the models. Moreover, for the standard model, the first relation is empty since the color sector is vectorial.
\par
We derived these relations for complex representations of complex algebras, but their general structure remains unchanged in the real case, since unitary elements of real and quaternionic matrices are traceless as well as their cubes. We still end up with three types of constraints: some rigid ones on the matrix of multiplicities, some linear equations for the $U(1)$ factors and a cubic relation for the "hypercharges". 
\par
There is an other reason to remove an unwanted $U(1)$. Since the Hilbert space must be equipped with a representation of an associative algebra, it is not possible, in general, to build a singlet under a given gauge group. To proceed, we introduce an additional summand $\cc$ in the algebra, so that this singlet sits in a representation of the associative algebra $\cc$. Then, we remove the corresponding $U(1)$ factor from the gauge group and we end up with a singlet under gauge transformations.  

\subsection{The fermionic representation}

Within the diagrammatic approach, to each couple $(i,j)$ of irreducible representations is associated a vertex $(i,j)$ of the diagram. This vertex corresponds to the Hilbert space $\hh_{ij}$ appearing in the decomposition of $\hh$. $\hh_{ij}$ is endowed with a bimodule structure given by $i$ (left action) and $\ov{j}$ (right action). The restriction of $\pi(g)\jj\pi(g)\jj^{-1}$ to $\hh_{ij}$ is 
$g_{i}\ot I_{|\mu_{ij}|}\ot \ov{g}_{j}$, so that vectors of $\hh_{ij}$ transform in the tensor product $i\ot\ov{j}$ of the corresponding representations $i$ and $j$ of the simple factors of gauge group, with multiplicity $|\mu_{ij}|$. If the two factors are distinct, the tensor product is irreducible as representation of $G$ and, for each simple factor it is just the fundamental one or its complex conjugate. Otherwise, the tensor product is not irreductible. For $SO(n)$ and $SP(n)$, the tensor product of two fundamental representations yields the adjoint one, the symmetric and traceless one and a singlet. For $SU(n)$ the situation is different, since we can tensorize the fundamental one and its complex conjugate, and we decompose the resulting product using Young boxes. In the analysis, we do not care about the unimodularity conditions. However, the latter are simply linear conditions imposed on the $U(1)$, and do not change the general strucure of the fermionic representation.
\par  
Moreover, to each vertex is associated a sign corresponding to the eigenvalue of $\chi$ on $\hh_{ij}$. Accordingly, fermions of $\hh_{ij}$ are chiral ones, right-handed if $\mu_{ij}>0$ and left-handed if $\mu_{ij}<0$. As a consequence of the symmetry of $\mu_{ij}$ and of the rules of table 1, fermions of $\hh_{ij}$ and $\hh_{ji}$ have the same chirality as well as those of $\hh_{ij}$ and $\hh_{\ov{ij}}$ and those of $\hh_{i\ov{j}}$ and  $\hh_{\ov{i}j}$.
\par
The charge  conjugation $\jj$ maps $\hh_{ij}$ onto $\hh_{ji}$, so that, if $i\neq j$, fermions of $\hh_{ji}$ are the antifermions of those of $\hh_{ij}$ and $\hh_{ij}$ does not contain Majorana particles. Otherwise, $\hh_{ii}$ is stable under charge conjuagation, and there may be Majorana fermions. In general, there always are Majorana fermions in $\hh_{ii}$, that transforms as the tensor product $i\ot\ov{i}$, except when the spectral triple is $S^{0}$-real.
\par
Many general properties of the fermionic sector, such as chirality, charge conjugation and transformation laws can be read on the diagram. As an example, let us try to build the $SU(5)$ grand unified model. To proceed, we need two representations of $SU(5)$: the $\ov{5}$ for right-handed fermions and the $10$ for left-handed ones. The $\ov{5}$ can be obtained with the algebra $M_{5}(\cc)\op\cc$ and the corresponding bimodule structure is given by the couples $(\ov{5},1)$ or $(\ov{5},\ov{1})$. At the gauge group level, we remove the unwanted U(1) with a unimodularity condition, as well as the one arising from $M_{5}(\cc)$. If we tensorize the $5$ by itself we get the 10 and the 15. Thus, if we remove the fermions sitting in the 15, it works perfectly well from the point of view of representation theory. In the basis composed by the representations $M_{5}(\cc)$, $\ov{M_{5}(\cc)}$ and $\cc$, the matrix of multiplicities reads, with $N_{f}$ generations,
$$
\mu=N_{f}\,\pp{0&-1&0\cr -1&0&1\cr 0&1&0\cr}.
$$
The corresponding diagram is given in figure 2.
\begin{figure}
\centering
\epsfig{file={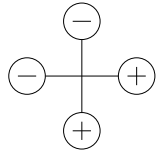},width=2cm}
\caption{Diagram for grand unified $SU(5)$ model}
\end{figure}
\par
To end this section, let us note that it is impossible to build the $SO(10)$ model out of a spectral triple. Indeed, it requires the 16 (spinorial representation of $SO(10)$) that cannot be obtained as a subrepresentation of a tensor product of the fundamental one or its complex conjugate. 

\subsection{The scalar fields and their representations}

To determine the Higgs fields and their transformation laws, we first have to compute explicitely the space of 1-forms. We recall that $\Delta$ is a 1-form and that the space of 1-forms $\Omega_{\dd}^{1}(\aa)$ is the bimodule generated by $\Delta$. Thus, if $(E_{ij}^{p})_{1\le p\leq n_{i}n_{j}}$ is a basis of $M_{n_{i}\times n_{j}}(\cc)$, the 1-forms
$$
\pi(E_{ii}^{p})\;\Delta\;\pi(E_{jj}^{q})
$$
span the space $\Omega_{\dd}^{1}(\aa)$.
\par
However, they are not, in general, linearly independent. To find a basis of $\Omega_{\dd}^{1}(\aa)$, we must have a closer look at the matrix structure of $\Delta$. To simplify our discussion, let us assume that we are dealing with a complex spectral triple. In this case, it is useful to choose $(E_{ij}^{p})_{1\leq p\leq n_{i}n_{j}}$ as  the canonical basis of $M_{n_{i}\times n_{j}}(\cc)$. We denote it by $E_{ij}^{ab}$, and the multiplication rule is simply
$$
E_{ij}^{ab}\,E_{kl}^{cd}=\delta_{jk}\delta_{bc}\,E_{il}^{ad}.
$$
As in section 4.2, we rewrite the matrix elements of $\Delta$ as
$$
\Delta_{ik}^{jl}=\delta_{kl}\,\mathop{\sum}\limits_{p}
E_{ij}^{ab}\ot M_{ij,k}^{ab}\ot I_{n_{k}}.
$$
Therefore,
$$
\pi(E_{ii}^{ab})\Delta\pi(E_{jj}^{cd})=\mathop{\sum}\limits_{k}
P_{ik}^{*}\lp E_{ij}^{ad}\ot M_{ij,k}^{bc}\ot I_{n_{k}}\rp P_{jk}.
$$
To construct a basis, let us consider the matrices $(M_{ij,k}^{ab})_{1\leq k\leq N}$ as a column vector $M_{ij}^{ab}$. The vectors $M_{ij}^{ab}$ span, when $a$ and $b$ vary, a vector space $V_{ij}$ whose dimension is denoted by $r_{ij}$. Out of the vectors $(M_{ij}^{ab})$, we choose a basis of $V_{ij}$ that we denote by $(M_{ij}^{p})_{1\leq p\leq r_{ij}}$. The components $M_{ij,k}^{p}$ of $M_{ij}^{p}$ yield a basis $\theta_{ij}^{ab,p}$ of $\Omega_{\dd}^{1}(\aa)$ defined by
$$
\theta_{ij}^{ab,p}=\mathop{\sum}\limits_{k}
P_{ik}^{*}\lp E_{ij}^{ab}\ot M_{ij,k}^{p}\ot I_{n_{k}}\rp P_{jk}.
$$
Accordingly, $\Omega_{1}^{\dd}(\aa)$ is a vector space of dimension $\sum_{ij} n_{i}r_{ij}n_{j},$ and it can be parameterized by complex matrices as 
$$
\Omega_{\dd}^{1}(\aa)=\la\mathop{\sum}\limits_{ijkp}
P_{ik}^{*}\lp \theta_{ij}^{p}\ot M_{ij,k}^{p}\ot I_{n_{k}}\rp P_{jk}
\;|\;\theta_{ij}^{p}\in M_{n_{i}\times n_{j}}(\cc)\ra.
$$
In the real case, it works in much the same way if $i$ and $j$ label irreducible representations of $\aa$. However, in the quaternionic case we must tensorize the elementary matrices with Pauli matrices. We also have some additional constraints between the matrices $\theta_{ij}$ such as $\theta_{\ov{ij}}=\ov{\theta}_{ij}$, where $\ov{i}$ denotes the representation complex conjugate of $i$ and $\ov{\theta}$ is either the complex or the $SU(2)$ conjugation.
\par
This applies in particular for the Higgs field $\Phi$ that can be parameterized by matrices $\phi_{ij}^{p}$ as
$$
\phi=\mathop{\sum}\limits_{ijkp}
P_{ik}^{*}\lp\phi_{ij}^{p}\ot M_{ij,k}^{p}\ot I_{n_{k}}\rp P_{jk}.
$$
Since $\Phi$ is hermitian, we have
$$
\lp\phi_{ij}^{p}\rp^{*}=\phi_{ji}^{p}.
$$
Under a gauge transformation, 
$$
\phi\longrightarrow \pi(g)\Phi\pi(g^{-1}),
$$
the matrix valued scalar fields $\phi_{ij}^{p}$ transform as
$$
\phi_{ij}^{p}\longrightarrow g_{i}\phi_{ij}^{p}g_{j}^{-1},
$$
where, as usual, $g_{i}$ denotes the image of $g$ through the irreducible representation labelled by $i$. Accordingly, the fields $\phi_{ij}^{p}$ transforms as the tensor product $i\ot\ov{j}$ of the two irreducible representations $i$ and $\ov{j}$.
\par
On the diagram, the fields $\phi_{ij}^{p}$ correspond to vertical links between the rows $i$ and $j$. Since such links cannot occur within the same row, the Higgs fields do not transform as the tensor product $i\ot\ov{i}$, but all other tensor products are allowed. In particular, we never end up with Higgs fields in the adjoint representation. Therefore, it is not possible to build the conventional $SU(5)$ grand unified model in our framework. Although we proved that it can work at the level of the fermionic representation, to break $SU(5)$ down to $SU(2)\times U(1)\times SU(3)$, we need some Higgs fields sitting in the adjoint representation of $SU(5)$.
\par
The Yukawa coupling between $\Phi$ and $\Psi$,
$$
\langle\Psi,\lp\Phi+\jj\Phi\jj^{-1}\rp\Psi\rangle,
$$  
can be expressed in terms of the fields $\phi_{ij}^{p}$ if we expand $\Psi$ over the tensor products we introduced in section 3.2. As a result, it appears the Yukawa couplings between the chiral fermions of the vertices $(i,k)$ and $(j,l)$ and the scalar fields $\phi_{ij}$ or $\ov{\phi}_{kl}$ correspond to the vertical or horizontal edges that can occur between the vertices $(i,k)$ and $(j,l)$.
\par
According to the spectral action principle, the Higgs potential is a linear combination of terms like
$$
\t\lp \Phi+\jj\Phi\jj^{-1}\rp^{n}.
$$
Using the explicit form of $\Phi$ in terms of the Higgs fields $\phi_{ij}^{q}$, the previous function can be expanded over monomials like 
$$
\t\lp\phi_{i_{1}i_{2}}^{p_{1}}...\phi_{i_{r}i_{1}}^{p_{r}}\rp\,\
\t\lp\phi_{j_{1}j_{2}}^{q_{1}}...\phi_{j_{s}j_{1}}^{q_{s}} \rp,
$$
with $r+s=n$. The coefficient of this monomial in the Higgs potential depends on the matrices $M_{ij,k}^{p}$. It corresponds canonically to a loop of length $n$ contained in the diagram, made of $r$ horizontal and $s$ vertical edges. When we are dealing with loops, we must be aware that we have actually two links between the vertices $(i,k)$ and $(j,l)$, one for $M_{ij,k}$ and the other for $M_{ji,k}$, even if, for simplicity, we represent only one. For instance, the sequence of vertices,
$$
(i,k)\rightarrow (j,k)\rightarrow (i,k),
$$ 
must be considered as a loop. We refer to this kind of loops, obtained by going forwards and backwards along a tree, as trivial loops. The previous example corresponds, at the level of the Higgs potential, to the monomial
$$
\t\lp\phi_{ij}\phi_{ji}\rp=\t\lp\phi_{ij}\phi_{ij}^{*}\rp,
$$ 
which is a mass term. In general, mass terms only corresponds to trivial loops. Non trivial ones, such as the sequence
$$
(i,k)\rightarrow(j,k)\rightarrow(j,l)\rightarrow(i,l)\rightarrow(i,k),
$$ 
with $i\neq j$ and $k\neq l$, can only occur for $n\geq4$, that is, for interacting terms. 

\section{Conclusion}

Starting with the notion of finite spectral triple derived from general axioms,  we give all possible noncommutative geometries on finite spaces. The latter are given, up to unitary equivalence, by a symmetric and non degenerate matrix of multiplicities $\mu$ with entries in $\zz$ and an operator $\Delta$. The underlying bimodule strucrure is encoded in the modules of the entries of the matrix of multiplicities whereas their signs correspond to the chirality. The matrix is symmetric if only if it corresponds to a real structure, and it follows from Poncar\'e duality that it is non degenerate. Finite spectral triples are classified and constructed using diagrams whose vertices correspond to their matrix of multiplicities and whose edges represent the structure of the Dirac operator.
\par
Finite spectral triples are also tools to build some Yang-Mills-Higgs models, 
and a general feature of the resulting models can be summarized as follows:
\begin{itemize}
\item
The gauge group is, up to some $U(1)$ factors, a direct product of the simple Lie groups $SO(n)$, $SU(n)$ and $SP(n)$.
\item
Fermions transform as tensor products of two fundamental representations (or their complex conjugate) of the simple factors. Multiplets of chiral fermions correspond to the vertices of a diagram labelled by couples of fundamental representation or their complex conjugate.
\item
The edges of the previous diagram correspond naturally to Yukawa couplings. The resulting Higgs fields also transform as tensor products of two fundamental representations (or their complex conjugate)of the simple factors of the gauge group, except those leading to the adjoint one.  
\item
The Higgs potential, computed with the spectral action principle, can be expanded over monomials that correspond to loops of the diagram.
\end{itemize}
This study confirms,  in a more general setting, some earlier results pertaining to the standard model \cite{asq}, its $U(1)$ extensions \cite{kra} and grand unified theories \cite{guni}. Furthermore, such a classification has also been worked out independently by M. Paschke and A. Sitarz in \cite{sita}, putting emphasis on Hopf algebra symmetries.

\vskip 0.5truecm
\noindent

{\Large\bf Aknowledgements}\\
It is a pleasure to thank B. Iochum and T. Sch\"ucker for their unvaluable help and advices as well as A. Sitarz and M. Paschke for useful remarks and comments.


\end{document}